\documentclass[epjCONF]{svjour}
\usepackage{amsmath}
\usepackage{graphics}
\usepackage[varg]{txfonts} 
\usepackage[latin1]{inputenc}

\session-title{Hot and Cold Baryonic Matter -- HCBM 2010}
\begin{document}
\title{Hybrid approaches to heavy ion collisions and future perspectives}
\author{Marlene Nahrgang\inst{1,2}\fnmsep\thanks{\email{nahrgang@fias.uni-frankfurt.de}}\and Christoph Herold\inst{1,2}\and Stefan Schramm\inst{1,2} \and Marcus Bleicher\inst{1,2} }
\institute{Institut f\"{u}r Theoretische Physik, Goethe-University Frankfurt\\ Max-von-Laue Str. 1, 60438 Frankfurt am Main, Germany  \and Frankfurt Institute for Advanced Studies (FIAS)\\ Ruth-Moufang-Str. 1, 60438 Frankfurt am Main, Germany}
\abstract{
We present the current status of hybrid approaches to describe heavy ion collisions and their future challenges and perspectives. First we present a hybrid model combining a Boltzmann transport model of hadronic degrees of freedom in the initial and final state with an optional hydrodynamic evolution during the dense and hot phase. Second, we present a recent extension of the hydrodynamical model to include fluctuations near the phase transition by coupling a chiral field to the hydrodynamic evolution.
} 
\maketitle
\section{Introduction}
\label{intro}
Heavy ion collisions are an excellent tool to study the properties of dense and hot nuclear matter \cite{Scherer:1999qq}. One of the most interesting features to explore in heavy ion collisions is the phase diagram of quantum chromodynamics (QCD). At high temperatures and large baryonic densities the degrees of freedom of strongly interacting matter are supposed to be partonic. There are indications that this state of matter, the quark-gluon plasma, was created in heavy ion collisions at RHIC and behaves as an almost ideal fluid \cite{Adams:2005dq,Back:2004je,Arsene:2004fa,Adcox:2004mh}. It is one of the major challenges to learn about the quark-gluon plasma (QGP) from the final hadronic state observed in the detectors. The phase transition of QCD is twofold: besides the transition from deconfined partonic to confined hadronic matter, the chiral symmetry gets spontaneously broken at lower temperatures while the symmetry is restored in the high temperature phase. Lattice QCD studies have revealed that these two aspects form one crossover transition at least for vanishing baryochemical potential at critical temperatures given between $T_c=[145,165]$ MeV \cite{Borsanyi:2010bp,Soldner:2010xk}. Studies of effective low energy models of QCD show that the phase transition is a first order transition at high baryon densities and low temperatures \cite{Scavenius:2000qd}. Consequently the first order phase transition line terminates at a critical end point (CEP) in the $T-\mu_B$ plane of the phase diagram.

A broad variety of experimental signatures is proposed to study the QCD phase diagram in experiment. One very famous signature is the 'horn' structure in the excitation function for the kaon to pion ratio \cite{:2007fe}. It was attributed to the onset of deconfinement at SPS energies \cite{Gazdzicki:1997hm}. Other proposed signatures cover aspects of the chiral phase transition, a suggested critical point \cite{Stephanov:1998dy,Stephanov:1999zu} and the first order phase transition \cite{Mishustin:1998eq}. It is, however, not clear how much of a potential signal is developed in a realistic scenario of a heavy ion collision. The fast dynamics present during the creation and expansion of the fireball of strongly interacting matter makes it difficult to defer results from thermodynamic calculations directly. A thorough theoretical understanding of phase transitions in the environment of a heavy ion collision is necessary to make profound predictions for the running low energy RHIC program and the upcoming FAIR project.

We present results from a hybrid model approach to heavy ion collisions \cite{Petersen:2008dd,Andrade:2006yh,Teaney:2001av,Hirano:2007gc} and give an outlook on chiral fluid dynamic models which allow for the explicit propagation of critical fluctuations through the phase transition.

\section{The hybrid approach}
The hybrid approach describes a heavy ion collision in three steps, the inital state, the hot and dense stage with the phase transition and the final interactions after hadronization. By fixing the inital state and the freeze-out of particles from the hydrodynamically propagated densities one can focus on the effects of the phase transition and viscosity in the intermediate state. The model presented here is an extention of the standard transport UrQMD model by the optional use of an intermediate hydrodynamic evolution \cite{Petersen:2008dd}. This hybrid approach is realized in the latest version 3.3 of UrQMD and can be downloaded from www.urqmd.org.
\subsection{The initial state}
The inital collisions are obtained from the Ultra-relativistic Quantum Molecular Dynamics model \cite{Bass:1998ca,Bleicher:1999xi}. When the Lorentz-contracted nuclei have passed through each other all nucleons have potentially interacted at least once. This is the earliest time to assume local thermalization, which is needed for the validity of a hydrodynamic description of the expansion of the created matter. The transition time to the hydrodynamic evolution is
\begin{equation}
 t_{\rm start}=\frac{2R}{\sqrt{\gamma_{\rm cm}^2-1}}\, ,
\end{equation}
where $R$ is the radius of the nuclei and $\gamma_{\rm cm}$ is the Lorentz gamma factor of the two colliding nuclei in the center of mass frame. To map the particles from the UrQMD to the hydrodynamic fields they are described by Gaussian distributions with a width of $\sigma=1$fm. The net-baryon density, the energy density and the initial velocity profiles are then transformed onto a $3$-dimensional space-grid to initialize the hydrodynamic evolution. Thereby the event-by-event fluctuations in the initial conditions are taken into account. The spectators are not considered in the hydrodynamic evolution but propagated further in the cascade.

\subsection{The equation of state in the hydrodynamic evolution}
The most important input to the hydrodynamic expansion of the system besides the initial conditions is the equation of state. It strongly influences the dynamics of the systems and naturally describes the phase transition in local equilibrium. In the framework of the hybrid model different equations of state can be implemented and their influence directly seen as the inital and final state remain unchanged. The hadron gas equation of state, for example, allows for a direct comparision of the underlying dynamics of the hydrodynamic treatment versus the transport sitmulation of the hot and dense region, since it includes the same degrees of freedom as in UrQMD and does not exhibit a phase transition. It can be used for baseline calculations to explore the influence of vanishing viscosity and local thermalisation. Other available equations of state are the Bag Model and the chiral equation of state. The Bag Model equation of state has a strong first order phase transition for all baryochemical potentials with a large latent heat. The chiral equation of state is obtained from a chiral hadronic Lagrangian that includes all baryons from the lowest flavor-SU($3$) octet and the multiplets of scalar, pseudo-scalar, vector and axial-vector mesons. Recently a deconfinement equation of state has been derived by including partonic degrees of freedom in the chiral Lagrangian \cite{Dexheimer:2009hi}, very much in the same manner as the Polyakov-loop extention of quark meson models. It has been successfully implemented to include investigations on the confinement-deconfinement transition to the framework of the hybrid model \cite{Steinheimer:2009nn}.

The hydrodynamic expansion is performed by a full ($3+1$)-dimensional SHASTA code \cite{Rischke:1995ir,Rischke:1995mt} solving for energy and momentum conservation of an ideal fluid. 

\subsection{Freeze-out}
When the system has expanded and diluted to a point where the hydrodynamical prescription becomes questionable and local equilibration can not be assumed any longer, a transition to a Boltzmann transport description is performed. In order to transfer the hydrodynamic densities to particles the Cooper-Frye freeze-out equation
\begin{equation}
 E\frac{{\rm d}N}{{\rm d}^3p}=\int_\sigma{\rm d}\sigma_\mu p^\mu f(p^\mu u_\mu,T,x)\, ,
\end{equation}
is employed. It connects the boosted phase-space distribution $f(x,p)$ and the momentum-space distribution ${\rm d}N/{\rm d}^3p$ along the hypersurface $\sigma$. For the present calculations a gradual transition is chosen, which corresponds approximately to an iso-eigentime transition. Transverse slices of thickness $\Delta z=0.2$fm are frozen out when the energy density $\epsilon$ in each cell of that slice has dropped below a critical energy density $\epsilon_{\rm crit}$.  It is important to note that for the correct transition the degrees of freedom of both sides, the hydrodynamic and the UrQMD \cite{Bravina:1998it}, need to be the same. Therefore, the last step in the hydrodynamic evolution is performed with an equation of state of a hadronic gas. The Monte Carlo sampling implementation of the Cooper-Frye equation conserves energy,baryon number, electric charge and strangeness in each single event. 

\section{Results from the hybrid model}
We present results from the hybrid approach to heavy ion collisions, particle multiplicities and ratios of strange particles to pions. 

\subsection{Particle multiplicities and mean transverse mass excitation function}
\begin{figure}
\resizebox{0.99\columnwidth}{!}{ \includegraphics{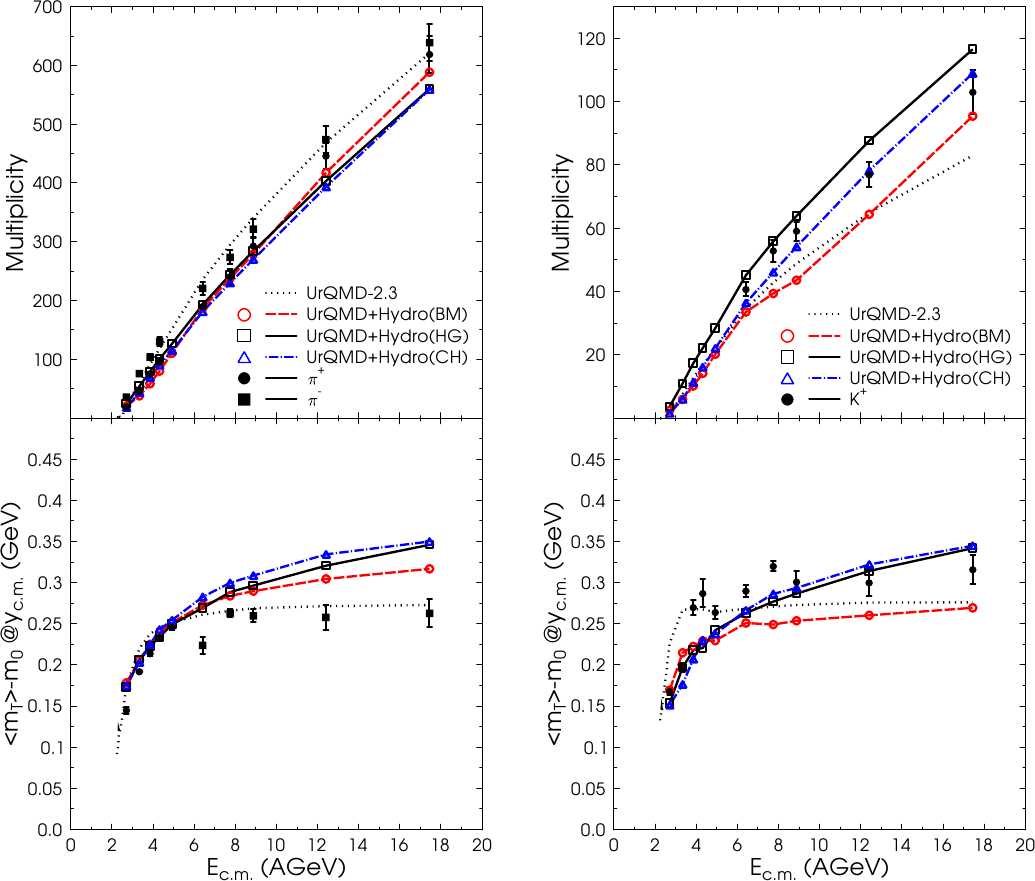}}
 \caption{Pion (left) and kaon (right) multiplicities (top) and mean transverse mass excitation function for central ($b<3.4$fm) Au+Au/Pb+Pb collisions. The experimental data (full black symbols) are compared to hybrdi model calculations with different equations of state (lines with symbols) and pure UrQMD 2.3 calculations (dashed line) \cite{Adams:2005dq,Adcox:2004mh,Afanasiev:2002mx,Ahle:1999uy,Klay:2003zf,Mitrovski:2006js}}
 \label{fig:multi}       
 \end{figure}

To study the dependence of the equation of state, viscous effects and nonequilibrium dynamics the hybrid model results can be compared to pure UrQMD 2.3 calculations. In Fig. (\ref{fig:multi}) the multiplicities (top) and the mean transverse mass excitation function (bottom) are shown for pions (left) and kaons (right) compared to data from most central Au+Au/ Pb+Pb collisions at intermediate center-of-mass energies $\sqrt{s}=2-18A$ GeV.
While the yield of pions is reduced in the hybrid model calculation compared to the transport calculation due to entropy conservation in the hydrodynamic expansion, as seen in the upper left plot of Fig. (\ref{fig:multi}), it does only show a weak dependance on the equation of state. In the kaon multiplicities, upper right plot of Fig. (\ref{fig:multi}), one clearly sees that the UrQMD 2.3 underpredicts the kaon yields in the whole energy range. In the Cooper-Frye freeze-out prescription the strange particles are produced according to their thermal distribution which leads to an enhanced yield in the hybrid model. Except for the Bag Model equation of state one sees a clear excess over the pure transport calculation, and a splitting between different equations of state.
Since the transverse mass spectra are (Fig. (\ref{fig:multi}, bottom) more sensitive to the pressure in the transverse plan it shows a more pronounced dependence on the equation of state. For both pions and kaons the chiral and the hadron gas equation of state give similar results. Only the Bag Model equation of state yields lower pion mean transverse masses at higher energies as it is expected for a first order phase transition. Again in the Bag Model equation of state the values of the mean transverse masses for kaons, lower right plot of Fig. (\ref{fig:multi}), are even below the nonequilibrium dynamics of the transport model.

\subsection{The kaon to pion ratio}
\begin{figure}
\resizebox{0.99\columnwidth}{!}{ \includegraphics{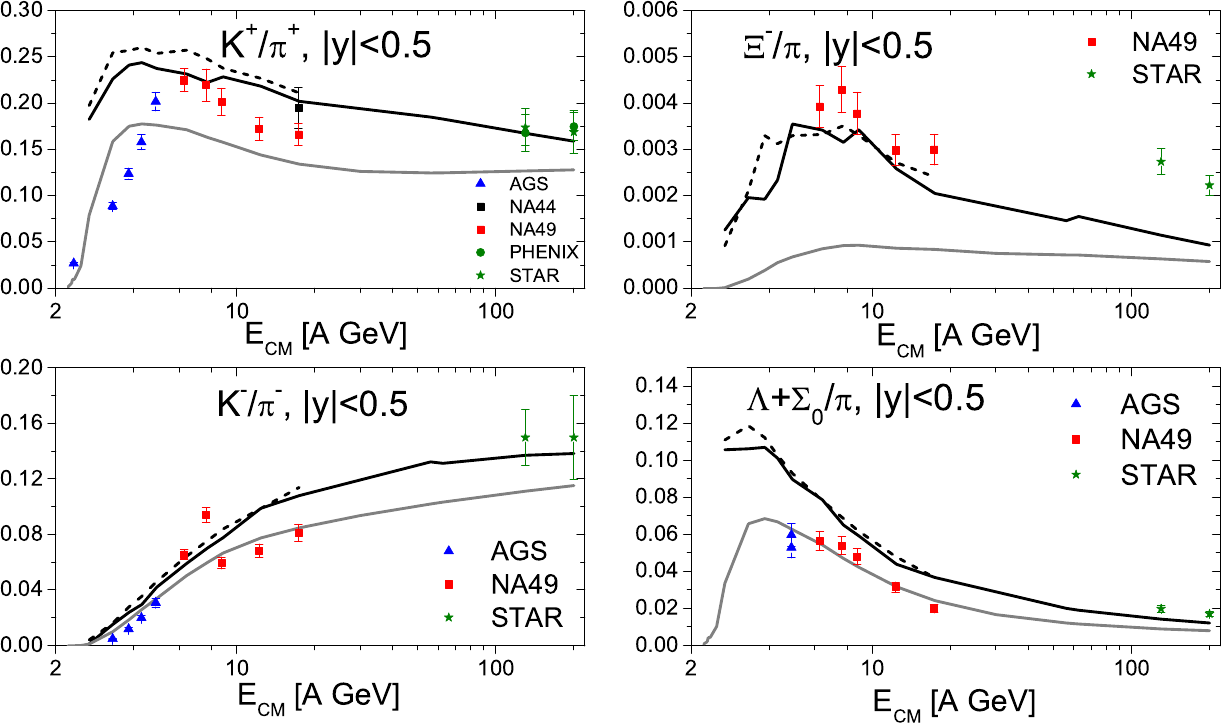}}
 \caption{Excitation function for kaon to pion ratios (left) and strange-baryon to pion ratios (right) for most central Au+Au/Pb+Pb collisions in the energy range $\sqrt{s}=2-200A$ GeV. The hybrid model calculations are the solid and dashed black line for two different freeze-out criteria. UrQMD 2.3 results and experimental data \cite{Adams:2005dq,Adcox:2004mh,Afanasiev:2002mx,Ahle:1999uy,Klay:2003zf,Mitrovski:2006js} are shown for comparision.}
 \label{fig:ktopi}       
 \end{figure}

The strangeness production is enhanced in the hybrid model due to the local thermal equilibrium asssumed for the hydrodynamic expansion \cite{Petersen:2009zi}. We can, therefore, expect to obtain different results for the kaon to pion and strange baryon to pion ratios. These are shown for central Au+Au/Pb+Pb collisions at center-of-mass energies $\sqrt{s}=2-200A$ GeV in Fig. (\ref{fig:ktopi}). There are two different freeze-out criteria of $\epsilon_{\rm crit}=4\epsilon_0$ (solid line) and $\epsilon_{\rm crit}=5\epsilon_0$ (dashed line). The gray line is a pure UrQMD 2.3 calculation that underestimates the data in all four cases for higher energies. This is due to an over-production of pion yields compared to the strange particle production. As expected the hybrid model gives a better description of the data for energies above $\sqrt{s}\approx 5$ GeV. At lower energies the assumption of local thermal equilibrium might not be justified and nonequilibrium propagation seems to be necessary. 

\section{Critical fluctuations in chiral hydrodynamics}
While the hybrid model gives an excellent framework to study the effects of an (local) equilibrium versus a full nonequlibrium transport propagation, it is not possible to propagate fluctuations directly in the hydrodynamic evolution.
In models of chiral hydrodynamics \cite{Mishustin:1998yc,Paech:2003fe} critical fluctuations at the phase transition can be studied directly and in non-equilibrium. For this purpose a field theoretical model that exhibits a phase transition is coupled to a background hydrodynamic evolution to simulate the expansion and cooling of the fireball in the course of a heavy ion collision. An effective model of QCD that exhibits a chiral phase transition and a critical point is the quark meson model with constituent quarks \cite{Scavenius:2000qd}
\begin{eqnarray}
{\cal L}=\bar{q}\left[i\gamma^\mu\partial_\mu-g\left(\sigma+i\gamma_5{\rm\textbf{$\tau$}}{\rm \textbf{$\pi$}}\right)\right]q 
  + \frac{1}{2}(\partial_\mu\sigma\partial^\mu\sigma)\\+
 \frac{1}{2}(\partial_\mu{\rm\textbf{$\pi$}}\partial^\mu{\rm\textbf{$\pi$}})
- U(\sigma, {\rm\textbf{$\pi$}})\nonumber \, ,
\label{eq:LGML}
\end{eqnarray}
with the constituent quark field $q=\left(u,d\right)$, the coupling $g$ between the quarks and the chiral fields $(\sigma,{\rm\textbf{$\pi$}})$. The potential is given by
\begin{equation}
U\left(\sigma, {\rm \textbf{$\pi$}}\right)=\frac{\lambda^2}{4}\left(\sigma^2+{\rm \textbf{$\pi$}}^2-\nu^2\right)^2-h_q\sigma-U_0\, ,
\label{eq:Uchi}
\end{equation} 
where the vacuum expectation of the sigma field is $\langle\sigma\rangle=f_\pi=93$~MeV and of the pion field is $\langle{\rm\textbf{$\pi$}}\rangle=0$. $h_q\sigma=f_\pi m_\pi^2$ with $m_\pi=138$~MeV is the explicit symmetry breaking term. Then $\nu^2=f_\pi^2-m\pi^2/\lambda^2$. $\lambda^2=20$ yields a sigma mass $m_\sigma^2=2\lambda^2 f_\pi^2 + m_\pi^2\approx 604$~MeV. In order to have zero potential energy in the ground state a term $U_0=m_\pi^4/(4\lambda^2)-f_\pi^2 m_\pi^2$ is added.

\subsection{The field equation of motion}
The order parameter of chiral symmetry breaking is the sigma field. From thermodynamic considerations it is expected to fluctuate largely at the critical point. Our purpose is, however, to propagate the chiral fields in nonequilibrium and study the effects of the phase transition. It is expected that critical fluctuations at the critical point are diminished while nonequilibrium fluctuations at the first order phase transition are enhanced. We will in the following concentrate on the evolution of the sigma field and neglect fluctuations of the pion fields around their vacuum expectation value. The sigma field interacts with the quarks and antiquarks, which we assume to be in local thermal equilibrium. Due to this interaction the sigma field is damped and according to Langevin dynamics gets random kicks from the heat bath.
The classical equation of motion for the chiral fields is 
\begin{equation}
 \partial_\mu\partial^\mu\sigma+\frac{\delta U}{\delta\sigma}+2g^2d_q\sigma\int\frac{{\rm d}^3p}{(2\pi)^3}\frac{1}{E}n_{\rm F}(E) +\eta\partial_t\sigma=\xi\, ,
\label{eq:eomsigma}
\end{equation}
where $n_{\rm F}(p)$ is the Fermi-Dirac distribution, $d_q=12$ the degeneracy factor and $E=\sqrt{p^2+g^2\sigma^2}$ the energy of the quarks. The mass of the quarks is dynamically generated by a nonzero value of the sigma field.
For the damping coefficient we take the value $\eta=2.2/fm$ \cite{Biro:1997va}. Then the noise term is, in Markovian approximation,
\begin{eqnarray}
 \langle \xi(t)\rangle&=&0\, ,\\
 \langle\xi(t)\xi(t')\rangle&=&\frac{2T}{V}\eta\delta(t-t')\; .
\end{eqnarray}

For the hydrodynamic evolution of the quark fluid the equation of state is needed. The pressure can be obtained from the thermodynamic relations in mean-field approximation
\begin{equation} 
V_{{\rm eff}}(\sigma, \vec{\pi},T)=-2d_q T \int\frac{{\rm d}^3p}{(2\pi)^3}\log(1+{\rm e}^{-\frac{E}{T}}) + U\left(\sigma, \vec{\pi}\right)\; .
\end{equation}
The strength of the phase transition at $\mu_B=0$ can be tuned by changing the strength of the coupling $g$. It is a crossover for small coupling and a first order phase transition for larger couplings, we use  $g=5.5$. Here, the effective potential has two degenerate minima allowing for the study of nucleation and spinodal decomposition \cite{Mishustin:1998eq,Chomaz:2003dz}. For an intermediate coupling of $g=3.63$ a critical point with one very flat minimum can be found.
Energy density and pressure are related by the thermodynamic relations
\begin{eqnarray}
  p(\sigma, \vec{\pi},T)&=& -V_{\rm eff}(\sigma,T)+U(\sigma)\; ,\\
  e(\sigma, \vec{\pi},T)&=& T\frac{\partial p(\sigma,T)}{\partial T}-p(\sigma,T)\; .
\end{eqnarray}
Besides the limited degrees of freedom in this model, the main difference to the chiral equation of state used in the hybrid model calculations above is that the field value is not the minimum of the thermodynamic potential. It is explicitely propagated and, thus, the energy density and the pressure explicitely depend on the actual value of the sigma field.
In the numerical simulations below the equation of motion of the sigma field (\ref{eq:eomsigma}) is solved by a staggered leap-frog algorithm. Initially we assume the sigma field to be in equilibrium with the fluid in order to be sensitive to the effects of the phase transition only.

\subsection{The hdyrodynamic expansion}
The equations of relativistic hydrodynamics of energy and momentum conservation are are
\begin{equation}
\partial_\mu (T_{\rm{fluid}}^{\mu\nu}+T_{\rm{field}}^{\mu\nu})=0
\label{eq:fluidT}
\end{equation}
where $T_{\rm{fluid}}$ is the energy-momentum tensor of an ideal fluid. The source term reflects the energy dissipation from the field into the fluid.
By using the explicit form of the equation of motion (\ref{eq:eomsigma}) we derive for the source term
\begin{equation}
S^\nu=-\partial_\mu T_{\rm field}^{\mu\nu}=-(-g\langle\bar q q\rangle_\sigma-\eta\partial_t\sigma+\xi)\partial^\nu\sigma
\label{eq:sourceterm}
\end{equation}
The inclusion of the source term guarantees energy and momentum conservation of the entire system.
The fluid dynamic part is again propagated by a SHASTA code. The energy density is initiated as the equilibrium energy density at $T=160{\rm MeV}$, ellipsoidal in x-y-plane and homogenous in z-direction, smoothed by a Woods Saxon type distribution.

\section{Numerical results}
The crucial quantity to look at is the intensity of the sigma fluctuations. It is given by
\begin{equation}
 \frac{{\rm d}N_\sigma}{{\rm d}^3k}=\frac{a_k^\dagger a_k}{(2\pi)^3 2\omega_k}=\frac{1}{(2\pi)^3 2\omega_k}{(\omega_k^2|\sigma_k|^2+|\dot\sigma_k|^2)}
\label{eq:numsig}
\end{equation}
This quantity can be considered as the number of sigma particles produced from the excited sigma field, once the nonlinearities can be neglected in the equation of motion of the sigma field.
In order to take into account the mass change of the sigma particle in a hot environment the average temperature $T_{\rm av}$ in the hot region is calculated. With the sigma mass
\begin{equation} 
 m_{\sigma}=\sqrt{\partial^2 V_{\rm eff}(\sigma,T_{\rm av})/\partial\sigma^2|_{\sigma=\sigma_{\rm eq}}}
\end{equation}
the energy of the modes is 
\begin{equation}
 \omega_k=\sqrt{|k|^2+m_{\sigma}^2}\, .
\end{equation}
\begin{figure}
\resizebox{0.99\columnwidth}{!}{ \includegraphics{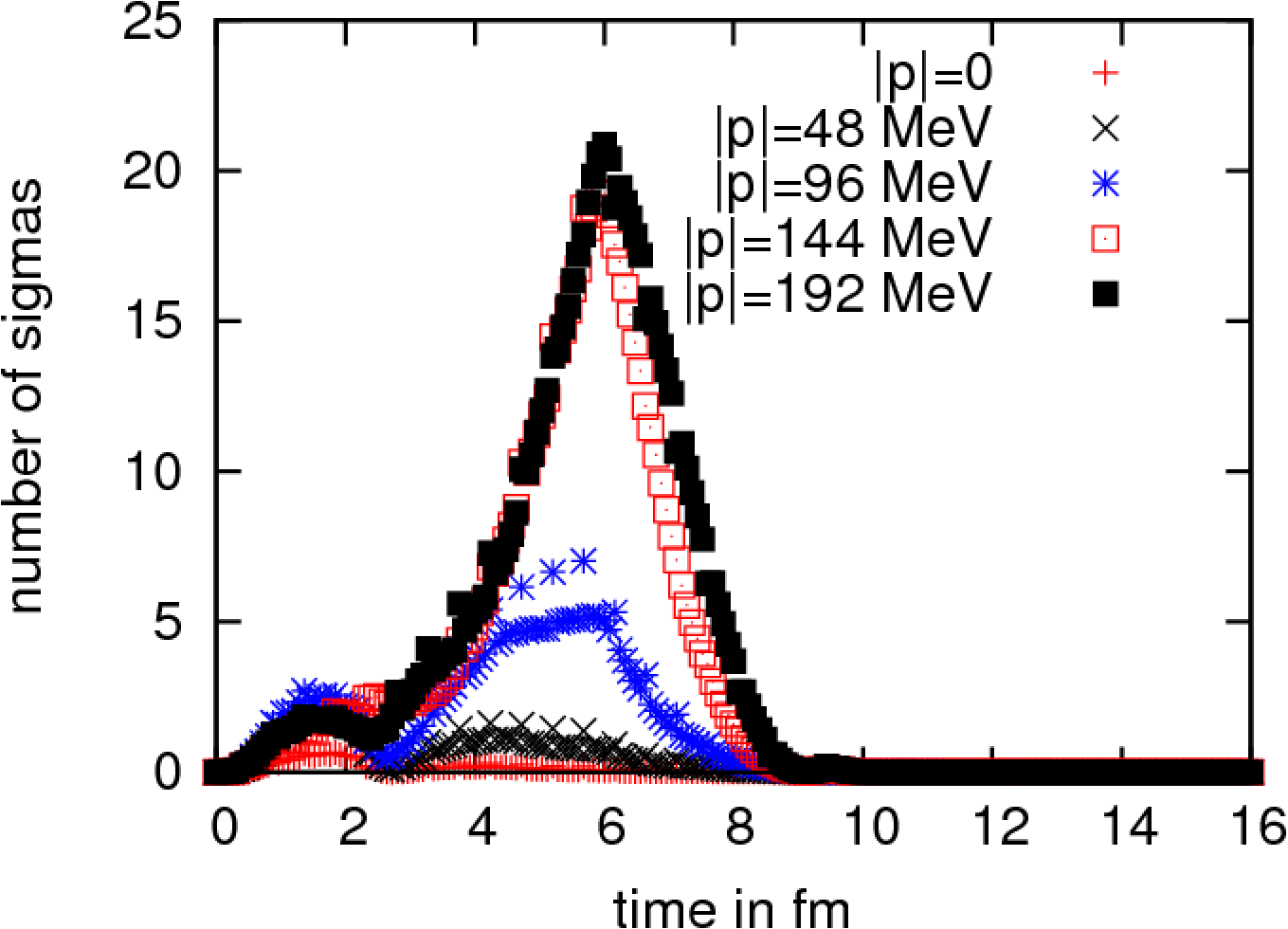}}
\caption{\label{fig:signumfo}Time evolution of the intensity of sigma fluctuations at a first order phase transition.}
\end{figure}
\begin{figure}
\resizebox{0.99\columnwidth}{!}{ \includegraphics{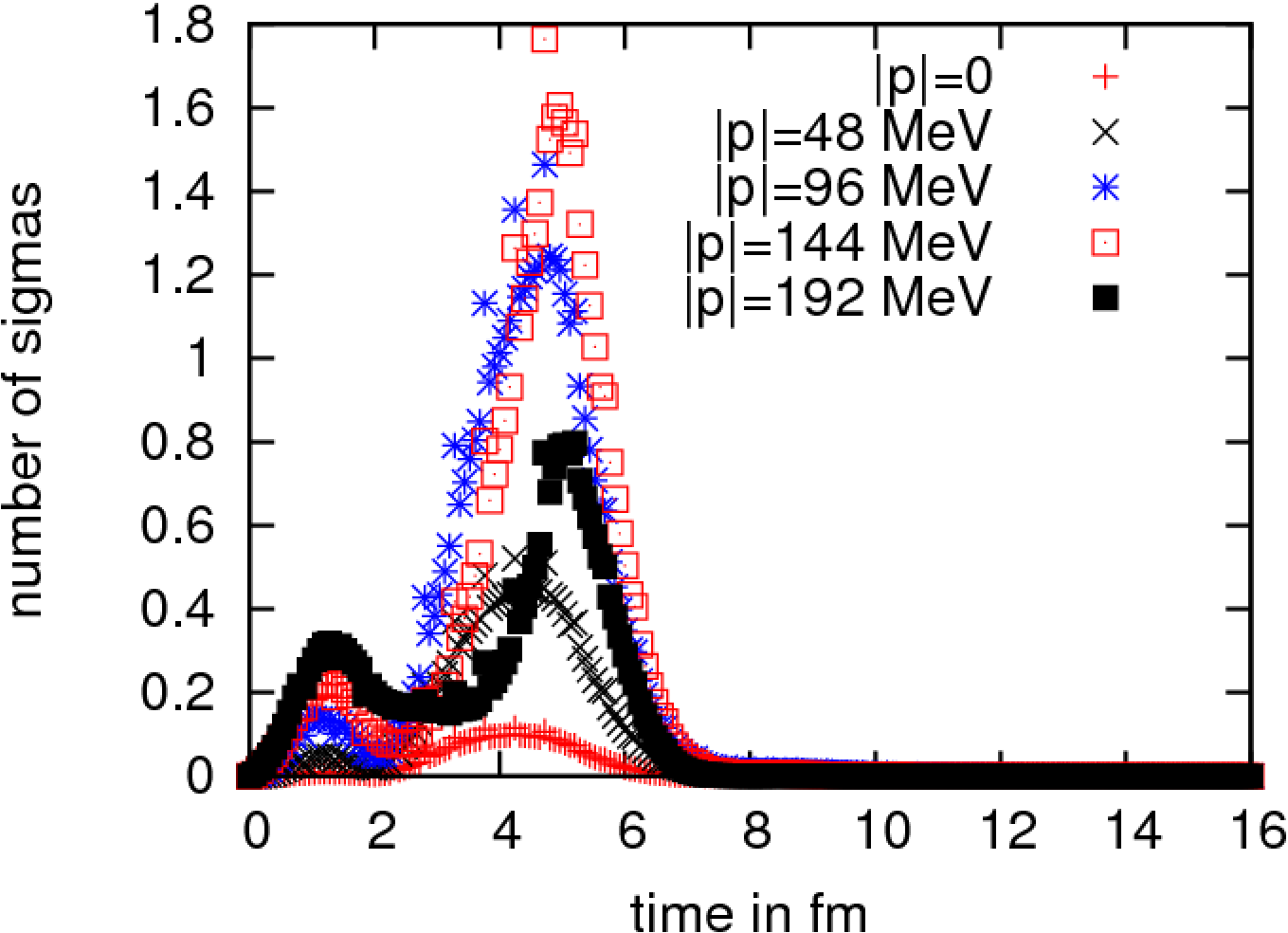}}
\caption{\label{fig:signumcp}Time evolution of the intensity of sigma fluctuations at a critical point.}
\end{figure}
The time evolution of (\ref{eq:numsig}) is shown in Fig. (\ref{fig:signumfo}) for a first order phase transition and in Fig. (\ref{fig:signumcp}) for a critical point. In a first order phase transition the intensity of sigma fluctuations between $4$ and $11$ fm is much larger than at a critical point. This is due to nonequilibrium effects at the first order phase transition where the effective potential has to two degenerate minima. Since we initialized the sigma field in equilibrium the increase in the sigma fluctuations is due to the phase transition. In the beginning of the evolution the sigma field is in the minimum which is stable above the critical temperatures $T>T_C$. It becomes the unstable minimum for temperatures below the critical temperature $T<T_C$. The barrier, which separates the two minima, however, prevents the sigma field from relaxing to the new equilibrium value. The damping to forces it further to remain in the local but unstable minimum. This effect is known as supercooling. The noise term turnes out to be too small to make nucleation an effective mechanism for relaxation. It is more likely that the field will remain in its unstable configuration almost until the barrier disappears and it can relax via spinodal decomposition \cite{Chomaz:2003dz}.

\section{Summary}
We have presented results from hybrid models to describe heavy ion collisions. The fully integrated Boltzmann and ($3+1$)-hydrodynamic model is well established to test the effects of a nonequilibrium versus a (local) equilibrium evolution and of viscosity. By switching between different equations of state and comparing to the original UrQMD 2.3 transport model the effects of a the phase transition have been studied. It is interesting to mention that the assumption of local thermal equilibrium in the hybrid model gives a good description of the 'horn' structure in $K^{+}/\pi^+$ ratios without invoking a phase transition.
The implementation of a chiral fluid dynamic model, which combines the explicit propagation of the order parameter of the chiral phase transition and a fluid dynamic expansion of a quark fluid, enables us to study the phase transition in heavy ion collisions beyond the equation of state. We have included damping and noise term in the nonequilibrium dynamics of the chiral field in order to study the enhancement of the fluctuations at the phase transition and the relaxation to equilibrium. We have found that the intensity of fluctuations of the sigma field is significantly larger at the first order phase transition than at the critical point where large fluctuations in equilibrium would be expected.

\section*{Acknowledgements}
This work was supported by the Hessian LOEWE initiative through the Helmholtz International Center for FAIR (HIC for FAIR). The computational resources
were provided by the Frankfurt Center for Scientific Computing (CSC). The authors thank Igor Mishustin and Hannah Peterson for fruitful and interesting discussions.

\end{document}